%% file: ANA-HMBS-2024-35-PAPER.tex
\documentclass[PAPER, atlasdraft=false, cernpreprint, USenglish, texmf, orcidlogo]{atlasdoc}
\usepackage{atlaspackage}
\usepackage{comment}

\usepackage{atlasbiblatex}

\usepackage{atlasphysics}
\usepackage{booktabs} %
\usepackage{siunitx}  %
\usepackage{amsmath}  %
\graphicspath{{logos/}{figures/}}

\AtlasTitle{Search for decays of the Higgs boson into pair-produced pseudoscalar particles decaying into $\tau^+\tau^-\tau^+\tau^-$ using $pp$ collisions at $\sqrt{s}=13$ TeV with the ATLAS detector}
\AtlasVersion{3.0}

\AtlasAbstract{%
A search for a pair of low-mass pseudoscalars $a$ that promptly decay into $\tau$-leptons is presented using 140~fb$^{-1}$ of proton--proton collision data at  $13$ TeV centre-of-mass energy recorded with the ATLAS detector at the Large Hadron Collider. The result is used to place constraints on exotic decays of the Higgs boson into four $\tau$-leptons, $H\to aa\to \tau^+\tau^-\tau^+\tau^-$.
This search focuses on events with either one or two $\tau$-leptons decaying into hadrons and neutrinos, and the remaining three or two $\tau$-leptons decaying into either electron or muon and neutrinos.
No significant excess is observed above the expected Standard Model background and upper limits at the 95\% confidence level on $\mathcal{B}(H\rightarrow aa\rightarrow \tau^+\tau^-\tau^+\tau^-)$ are set ranging from 0.06 to 0.23, depending on the mass $m_a$ ranging from 15 to 60 GeV.
}

\AtlasRefCode{HMBS-2024-35}

\PreprintIdNumber{CERN-EP-2026-026}

\arXivId{2603.08323}

\AtlasJournal{New Journal of Physics}
\AtlasJournalRef{New J. Phys. 28 (2026) 063801}
\AtlasDOI{10.1088/1367-2630/ae78f7}

\AtlasCoverEgroupAnalysisTeam{atlas-hmbs-2024-35-analysis-team@cern.ch}

\hypersetup{pdftitle={ATLAS document},pdfauthor={The ATLAS Collaboration}}

\begin{document}

\maketitle

Beyond-the-Standard-Model (BSM) pseudoscalar particles $a$ with masses below the electroweak scale were proposed to explain several open problems in particle physics, for example as dark matter interaction mediators~\cite{Silveira:1985rk,Pospelov:2007mp,Draper:2010ew,Ipek:2014gua,Martin:2014sxa,Clowe}, or in the context of models of neutral naturalness~\cite{Burdman:2006tz,Craig:2015pha,Curtin:2015fna}.
One interesting scenario posits new light pseudoscalars coupled to the Higgs field, which can drive electroweak phase transition at leading-order in perturbation theory~\cite{Profumo:2007wc,Blinov:2015sna}. If lighter than half the Higgs boson mass ($m_a < m_H/2$), these pseudoscalars can be produced in pairs in exotic decays of the Higgs boson~\cite{Curtin:2013fra}.
The narrow natural width of the Higgs boson predicted by the Standard Model ($\Gamma_H^{SM}=3.7$~\MeV)~\cite{ParticleDataGroup:2024cfk} implies that even a small BSM partial width for an exotic decay can result in a sizeable branching ratio. Together with the large number of Higgs bosons produced in the proton--proton ($pp$) collisions at $\sqrt{s}=13$~\TeV at the Large Hadron Collider (LHC)~\cite{Evans:2008zzb}, such light pseudoscalars $a$ could become detectable through $H\rightarrow aa$ decays, even for very feeble coupling constants~\cite{Shrock:1982kd,Strassler:2006im,Schabinger:2005ei,Patt:2006fw}.

A search for the production of a pair of new pseudoscalar particles promptly decaying into four $\tau$-leptons is presented. The result presented uses the full Run-2 collision data at a centre-of-mass energy $\sqrt{s}=13~\TeV$ recorded with the ATLAS detector. The data collected from 2015 to 2018 corresponds to an integrated luminosity of $140\,\text{fb}^{-1}$.
The search targets the process $H\to aa\to \tau^+\tau^-\tau^+\tau^-$ and probes the mass range $15\,\GeV<m_a<60\,\GeV$.
The sensitivity is limited below the lower mass value by the angular separation required for the final-state objects, while it is limited beyond the upper mass value by the kinematic threshold for on-shell Higgs boson decays, $m_H/2=62.5\,\GeV$.
This search complements a previous search by the ATLAS Collaboration that focused on the lower mass range $4\,\GeV<m_a<15\,\GeV$ where the $\tau$-leptons are Lorentz-boosted and overlap, requiring a different analysis strategy~\cite{HMBS-2024-25}.
Limits have also been set on this model by the CMS Collaboration~\cite{CMS-HIG-14-019,CMS-HIG-18-006}.
The ATLAS Collaboration has also performed searches for new pseudoscalar particles produced in exotic Higgs boson decays in the $b\bar{b}b\bar{b}$~\cite{HIGG-2016-01,HIGG-2017-05,HDBS-2018-47,HMBS-2024-26}, $b\bar{b}\tau^+\tau^-$~\cite{HIGG-2016-16}, $b\bar{b}\mu^+\mu^-$~\cite{HDBS-2021-03}, and $\ell^+\ell^-\ell^{\prime\,+}\ell^{\prime\,-}$($\ell$ = $e$ or $\mu$)~\cite{EXOT-2013-15,EXOT-2016-22} final states.
The $H\rightarrow aa$ searches can also be interpreted within the Two-Higgs-Doublet Model~\cite{Branco:2011iw} extended by a new light pseudoscalar singlet (2HDM+S) framework~\cite{a0,Curtin:2013fra}. The 2HDM+S model effectively captures the phenomenology of the Higgs sector in the Next-to-Minimal Supersymmetric Standard Model (NMSSM), which extends the Minimal Supersymmetric Standard Model (MSSM) by adding a singlet field~\cite{a0}. In this context, the $\tau^{+}\tau^{-}\tau^{+}\tau^{-}$ final state is expected to complement the existing searches, particularly for Type III models with large $\tan\beta$~\cite{Curtin:2013fra}, where the new pseudoscalar decays almost exclusively to $\tau$-leptons. This makes the four-$\tau$ final state particularly sensitive in the mass range probed in this analysis, which fills the gap between the existing low-mass ATLAS result~\cite{HMBS-2024-25} and the kinematic threshold of the $H\to aa$ decay. The resulting limits on $\mathcal{B}(H\to aa\to 4\tau)$ therefore constrain the mixing between the Higgs boson and the new pseudoscalar state and, more broadly, the viable parameter space of scenarios in which such a light pseudoscalar acts as a portal to a dark sector or drives a strong first-order electroweak phase transition.

\newcommand{\AtlasCoordFootnote}{%
ATLAS uses a right-handed coordinate system with its origin at the nominal interaction point (IP)
in the centre of the detector and the \(z\)-axis along the beam pipe.
The \(x\)-axis points from the IP to the centre of the LHC ring,
and the \(y\)-axis points upwards.
Polar coordinates \((r,\phi)\) are used in the transverse plane,
\(\phi\) being the azimuthal angle around the \(z\)-axis.
The pseudorapidity is defined in terms of the polar angle \(\theta\) as \(\eta = -\ln \tan(\theta/2)\) and is equal to the rapidity
$ y = \frac{1}{2} \ln \left( \frac{E + p_z}{E - p_z} \right) $ in the relativistic limit.
Angular distance is measured in units of \(\Delta R \equiv \sqrt{(\Delta y)^{2} + (\Delta\phi)^{2}}\).}

The ATLAS experiment~\cite{PERF-2007-01} at the LHC is a multipurpose particle detector
with a forward--backward symmetric cylindrical geometry and a near \(4\pi\) coverage in
solid angle.\footnote{\AtlasCoordFootnote}
It consists of an inner tracking detector surrounded by a thin superconducting solenoid
providing a \qty{2}{\tesla} axial magnetic field, electromagnetic and hadronic calorimeters, and a muon spectrometer.
The inner tracking detector covers the pseudorapidity range \(|\eta| < 2.5\).
It consists of silicon pixel, silicon microstrip, and transition radiation tracking detectors.
Lead/liquid-argon (LAr) sampling calorimeters provide electromagnetic (EM) energy measurements
with high granularity within the region \(|\eta|< 3.2\).
A steel/scintillator-tile hadronic calorimeter covers the central pseudorapidity range (\(|\eta| < 1.7\)).
The endcap and forward regions are instrumented with LAr calorimeters for EM and hadronic energy measurements up to \(|\eta| = 4.9\).
The muon spectrometer surrounds the calorimeters and is based on three large superconducting air-core toroidal magnets with eight coils each.
The field integral of the toroids ranges between \num{2.0} and \qty{6.0}{\tesla\metre} across most of the detector.
The muon spectrometer includes a system of precision tracking chambers up to \(|\eta| = 2.7\) and fast detectors for triggering up to \(|\eta| = 2.4\).
The luminosity is measured mainly by the LUCID--2~\cite{LUCID2} detector that is located close to the beampipe.
A two-level trigger system was used to select events~\cite{TRIG-2016-01}.
The first-level trigger is implemented in hardware and used a subset of the detector information to accept events at a rate close to \qty{100}{\kHz}.
This is followed by a software-based trigger that reduced the accepted rate of complete events to \qty{1.25}{\kHz} on average depending on the data-taking conditions.
A software suite~\cite{SOFT-2022-02} is used in data simulation, in the reconstruction and analysis of real and simulated data, in detector operations, and in the trigger and data acquisition systems of the experiment.

Objects are reconstructed by combining the information from different sub-detectors and using dedicated identification algorithms. Electrons are reconstructed by matching narrow clusters in the electromagnetic calorimeter with tracks in the inner detector. Electron candidates are required to satisfy $\pt > 7\,\GeV$ and $|\eta|<2.47$, where $\pt$ is the transverse momentum. Candidates in the transition region between the barrel and endcap calorimeters ($1.37 < |\eta| < 1.52)$ are excluded. All electrons are required to satisfy the $\emph{Tight}$ charge-identification algorithm~\cite{EGAM-2021-01} to reduce charge misidentification. The baseline electron selection is defined by the $\emph{Loose}$ working point of the ATLAS likelihood-based electron identification algorithm~\cite{EGAM-2021-01} where no isolation requirement is applied. A more exclusive electron selection requires the electron candidates to satisfy the $\emph{Medium}$ identification working point and the $\emph{Tight}$ track-based isolation~\cite{EGAM-2021-01}.

Muons are reconstructed by matching tracks in the muon spectrometer with tracks in the inner detector. Muon candidates are required to satisfy $\pt > 5\,\GeV$ and $|\eta|<2.5$. Additional conditions, corresponding to the $\emph{Medium}$ working point of the ATLAS muon identification algorithm~\cite{MUON-2018-03}, are imposed on the quality of the tracks to define the baseline muon selection criteria. The exclusive muon identification further requires the $\emph{Tight}$ track-based isolation~\cite{MUON-2018-03}.

Both electrons and muons are required to satisfy $|z_0\sin\theta| < 0.5\,\text{mm}$, where $z_0$ is the longitudinal impact parameter relative to the primary vertex (PV), which is the reconstructed vertex with the highest sum of squared transverse momenta of associated tracks. Electrons (muons) are require to further satisfy $|d_0|/\sigma_{d_0} < 5(7)$ where $d_0$ is the transverse impact parameter relative to the beam line and $\sigma_{d_0}$ is its uncertainty.
The $\tau$-lepton is long-lived, but it has a shorter lifetime compared to $b$-hadrons,
so the value of the impact parameter $|d_0|$ is optimised to maintain
good acceptance to $\tau$-lepton decays while still reducing contributions from
heavy-flavour hadron decays.

Jets are reconstructed using the anti-\kt\ algorithm~\cite{Cacciari:2008gp} as implemented in FastJet~\cite{Fastjet} with a jet radius parameter $R = 0.4$. The inputs to this algorithm are particle flow objects~\cite{PERF-2015-09}, which combine measurements from the ATLAS inner detector and calorimeters~\cite{PERF-2014-07} to improve the jet energy resolution and increase the jet reconstruction efficiency, especially at low jet \pt.
The jet energy scale is calibrated to particle-level using simulation and corrections obtained from in situ techniques~\cite{JETM-2018-05}. A multivariate jet vertex tagger is used to identify jets with $\pt<60~\GeV$ and $|\eta|<2.4$ as originating from the PV, suppressing jets from secondary collisions in the same bunch crossing of the event of interest (\pileup)~\cite{PERF-2014-03}. All jets are required to have $\pt > 20\,\GeV$ and $|\eta|<2.5$. Jets containing $b$-hadrons, referred to as $b$-jets, are identified using the DL1r algorithm~\cite{FTAG-2019-07}. The chosen working point has an efficiency of 85\% for selecting $b$-jets~\cite{PERF-2016-05} as estimated from $t\bar{t}$ simulation.

The reconstruction of $\tau_{\text{had}}$ candidates is seeded by jets formed using the anti-\kt\ algorithm with a radius parameter $R = 0.4$,
using topological clusters of calorimeter cells~\cite{PERF-2014-07} as inputs.
The $\tau_{\text{had}}$ candidates are required to satisfy $\pt>20\,\GeV$ and $|\eta|<2.5$, excluding the transition region $1.37 <|\eta| < 1.52$. The $\tau_{\text{had}}$ identification uses a recurrent neural network (RNN) algorithm~\cite{ATL-PHYS-PUB-2019-033}. All $\tau_{\text{had}}$ candidates are required to have exactly one or three charged-particle tracks associated with its seed jet. A baseline $\tau_{\text{had}}$ selection requires the candidate to satisfy the $\emph{VeryLoose}$ RNN working point, while a more exclusive $\tau_{\text{had}}$ selection requires the candidates to further satisfy the $\emph{Loose}$ RNN working point~\cite{ATL-PHYS-PUB-2019-033}. In addition, a dedicated multivariate electron veto is applied to $\tau_{\text{had}}$ candidates~\cite{ATL-PHYS-PUB-2015-045}.

The missing transverse momentum, with magnitude $\met$, is calculated as the negative vector sum of the transverse momenta of all reconstructed physics objects in events and a track-based soft term~\cite{JETM-2020-03}. The soft term is constructed from charged-particle tracks that are consistent with originating from the primary vertex but are not associated with any of the primary reconstructed objects.
This search uses events selected with a combination of triggers that require the presence of either a single lepton ($e$ or $\mu$) or a lepton pair with same/different flavours ($ee$, $\mu\mu$ or $e\mu$)~\cite{TRIG-2018-05, TRIG-2018-01}. These triggers varied throughout the data-taking period to manage the increasing luminosity and \pileup conditions.
The single-lepton triggers had thresholds ranging from 20~\GeV to 26~\GeV and the dilepton triggers had various combinations of \pt thresholds ranging from 10~\GeV to 24~\GeV for the two leptons, depending on the data-taking period and lepton flavour.
To ensure that the reconstructed leptons are associated with the trigger decision, offline leptons are geometrically matched to corresponding trigger-level objects from the specific trigger chain that accepted the event.

Two non-overlapping signal regions (SR) are used to capture different decay modes of the $\tau$-leptons. The first SR,\@ called $2\ell 2\tau_{\text{had}}$, targets decays of the kind $H\to aa\to (\tau_{\ell}\tau_{\text{had}})(\tau_{\ell^{\prime}}\tau_{\text{had}})$, where $\ell$ or $\ell^{\prime}$ denotes an electron or a muon. Events in the $2\ell 2\tau_{\text{had}}$ region are selected by requiring two same-charge electrons or muons, and two same-charge $\tau_{\text{had}}$. In addition, the total charge of the four selected leptons is required to be zero. The same-charge selection reduces the number of background events from Drell--Yan processes while being consistent with exotic Higgs boson decays into four $\tau$-leptons. The second SR,\@ called $3\ell 1\tau_{\text{had}}$, targets decays of the kind $H\to aa\to (\tau_{\ell}\tau_{\ell^{\prime}})(\tau_{\ell^{\prime\prime}}\tau_{\text{had}})$. Events in the $3\ell 1\tau_{\text{had}}$ region are selected by requiring three electrons or muons and one $\tau_{\text{had}}$. A charge-based selection is also used. The three electrons or muons are required to have a total charge $+1$ or $-1$. The four bodies are required to have a total charge of zero. To significantly reduce the large background from Drell--Yan production, a $Z$ boson veto is applied. Any opposite-charge electron or muon pair is required to have an invariant mass $m_{\ell\ell}$ outside the $Z$ boson mass window, satisfying $m_{\ell\ell} < 71\,\GeV$ or $m_{\ell\ell} > 111\,\GeV$.

In both signal regions, events are required to have zero $b$-jets to reduce the number of background events with top quarks. The leading and subleading leptons are required to have $\pT>$ 20 and 15 \GeV. These thresholds are chosen to be least 1 \GeV above the threshold of the loosest HLT selection in order to ensure a constant trigger efficiency above 98\% for all $a$-boson mass points. Finally, the visible mass ($m_{\text{vis}}$) of the four-body system in both SRs is required to satisfy $m_{\text{vis}} < 110\,\GeV$. The value of $110\,\GeV$, below the Higgs boson mass of $125\,\GeV$, accounts for the energy carried away by the neutrinos in $\tau$-lepton decays. After the event selection, there are only two relevant sources of background events.
The dominant background source are events with fake/non-prompt (FNP) leptons, estimated by a dedicated data-driven method~\cite{EGAM-2019-01} that is further explained below. The second source are diboson ($WZ$ and $ZZ$) events with prompt leptons that escape the $Z$ boson mass veto selection and is estimated with Monte Carlo (MC) simulations. The contribution from other processes, including $WW$, is found to be negligible.

The MC simulation of $ZZ$ and $WZ$ events is performed with the
\SHERPA[2.2.2]~\cite{Bothmann:2019yzt} generator,%
including off-shell effects and Higgs boson contributions where appropriate.
Fully leptonic final states are simulated using matrix elements at next-to-leading-order (NLO) accuracy in quantum chromodynamics (QCD) for up to one additional parton and at leading-order (LO) accuracy for up to three additional parton emissions.
Samples for the loop-induced processes \(gg \to VV\) (where $V=W,Z$) are simulated using LO-accurate matrix elements for up to one additional parton emission. The matrix element calculations are matched and merged with the \SHERPA parton shower based on Catani--Seymour dipole factorisation~\cite{Gleisberg:2008fv,Schumann:2007mg} using the MEPS@NLO prescription~\cite{Hoeche:2011fd,Hoeche:2012yf,Catani:2001cc,Hoeche:2009rj}.
The virtual QCD corrections are provided by the \OPENLOOPS library~\cite{Buccioni:2019sur,Cascioli:2011va,Denner:2016kdg}. The \NNPDF[3.0nnlo] set of parton distribution functions (PDFs) is used~\cite{Ball:2014uwa}, along with the dedicated set of tuned parton-shower parameters developed by the \SHERPA authors.

The signal $pp \to H \to aa \to \tau^{+}\tau^{-}\tau^{+}\tau^{-}$ is simulated using the gluon--gluon fusion (ggF) production mode at NLO accuracy in QCD using \POWHEGBOX[v2]~\cite{Nason:2004rx,Frixione:2007vw,Alioli:2010xd}, with the CT10 PDF set~\cite{Butterworth:2015oua}. The cross section is normalised to the Higgs boson inclusive production cross section 55.6 pb~\cite{deFlorian:2016spz} since the difference in the acceptance compared to other Higgs boson production modes is negligible. The decay of the Higgs boson and the further hadronisation is performed with \PYTHIA[8.245]~\cite{Sjostrand:2014zea} which is configured to use the EvtGen~\cite{Lange:2001uf} programme for the decay of $b$-hadrons, using the \AZNLO\ set of tuned parameters (\AZNLO tune)~\cite{STDM-2012-23}.

All generated events are simulated with the ATLAS detector simulation~\cite{SOFT-2010-01} based on \GEANT~\cite{Agostinelli:2002hh}, and reconstructed with the same software as the data. The signal samples are processed through a faster simulation where the full \GEANT simulation of the calorimeter response is replaced by a parameterisation of the shower shapes~\cite{SOFT-2010-01}.
The effect of \pileup was modelled by overlaying the simulated hard-scattering event with
inelastic $pp$ events generated with \PYTHIA[8.186]~\cite{Sjostrand:2007gs}
using the \NNPDF[2.3lo] set of parton distribution functions (PDF)~\cite{Ball:2012cx} and the
A3 set of tuned parameters~\cite{ATL-PHYS-PUB-2016-017}.

The main source of background is composed of fake/non-prompt leptons (from decays of heavy-flavour hadrons or jets faking leptons) from processes such as $WZ$, $Z/\gamma^*+\mathrm{jets}$ and $t\bar{t}$.
These processes can contribute both to prompt leptons from $W$ or $Z$ boson decays and fake/non-prompt leptons from $b$-quark decays and jets, which compose the mis-identified background of the analysis.
Despite the lifetime of the $\tau$-lepton, electrons and muons produced in $\tau\to \ell\nu_{\ell}\nu_{\tau}$ decays are considered prompt. The relaxed impact parameter selection used in the exclusive electron and muon selection criteria ensure good acceptance for most of these decays. Sources of fake/non-prompt electrons include mis-reconstructed jets and photon conversions. Non-prompt muons arise almost exclusively from the semileptonic decay of hadrons.
Finally, fake/non-prompt $\tau_{\text{had}}$ are mis-identified quark- or gluon-initiated jets.
Electrons mis-reconstructed as $\tau_{\text{had}}$ decays with one charged-particle track are estimated from MC. Leptons reconstructed with the wrong charge assignment are always considered fake/non-prompt.
This fake/non-prompt background is derived using a data-driven fake-factor method~\cite{EGAM-2019-01}, using
events in which at least one lepton fails to meet the exclusive selection criteria. %
A weight fake factor $f/(1-f)$, where $f$ is the fake rate -- the fraction of events for which the baseline lepton also satisfies the exclusive selection criterion -- is applied to each lepton failing to satisfy the exclusive selection criterion.
For events containing multiple objects failing to meet the exclusive selection criteria, a combinatoric weight is calculated to account for all possible combinations of fake objects while avoiding over-correction~\cite{EGAM-2019-01}.
The contribution of background processes with prompt leptons that fail to satisfy the exclusive selection criteria is removed using MC simulation but has a negligible impact on the final expected yield.

The fake factors are determined in two steps: they are first measured in a data control region enriched in $Z$+jets events, and this measure is then refined in FNP-enriched validation regions containing same-sign leptons.
In the first step, a $Z$+jets sample is selected by requiring $e^+e^-$ or $\mu^+\mu^-$ pairs,
where both leptons satisfy the exclusive selection criterion and $71 < m_{\ell\ell} < 111\,\GeV$, and an additional baseline lepton.
Events are required to satisfy $\met<30\,\GeV$ to reduce the contamination from $WZ$ events in this control region to a negligible level.
The measured fake factor for electrons and muons is extracted in bins of lepton $\pt$ and $\eta$. For taus, the fake factor is measured in bins of visible $\pt$ and as a function of the number of charged-particle tracks associated with the seed jet.
The measured fake factors range from approximately 0.3--0.6 for muons, 0.01--0.09 for electrons and
0.25--0.5 for $\tau_\text{had}$ candidates with one or three tracks.

The background processes that contribute to FNP leptons vary in proportions in the $Z$+jets regions and in the two SRs.\@
To assess the possible impact of such variation in composition, three validation regions (VR) are defined.
The validation regions are similar to the SRs but require exactly three objects in the final state. Two VRs, called $\ell e\tau_{\text{had}}$ and $\ell\mu\tau_{\text{had}}$, require the presence of two same-charge leptons (electrons or muons) and one $\tau_{\text{had}}$.
The $\ell e\tau_{\text{had}}$ and $\ell\mu\tau_{\text{had}}$ VRs are distinguished by the flavour of the subleading electron or muon. A third VR, called $1\ell 2\tau_{\text{had}}$, is formed by selecting events with two same-charge $\tau_{\text{had}}$ and one electron or muon.
In the VRs, similar to the SRs, the three-body visible mass is required to satisfy $m_{\text{vis}}^{\text{3-body}}<110\,\GeV$ to be kinematically compatible with the signal region topology, and the number of $b$-jets is required to be zero.

The $\ell e \tau_{\text{had}}$ ($\ell\mu\tau_{\text{had}}$) VR is designed such that the subleading electron (muon) is dominantly from FNP sources, and similarly, the subleading $\tau_{\text{had}}$ in the $2\tau_{\text{had}}1\ell$ VR is dominated by FNP as well.
The validation regions serve two purposes: constraining systematic uncertainties and providing a refined fake factor estimate.
A profile-likelihood fit to the observed yields in the VRs is performed to propagate the impact of experimental and modelling uncertainties into an uncertainty model for the fake factors. This fit includes dedicated modelling uncertainties evaluated by varying key selection criteria for the three types of objects in the final states, the $\tau_{had}$ seed jet width, the electron ambiguity (which discriminates against electrons from photon conversions), and the muon $|d_0|/\sigma_{d_0}$ selection.

\begin{figure}[htbp!]
\centering
\includegraphics[width=0.49\linewidth]{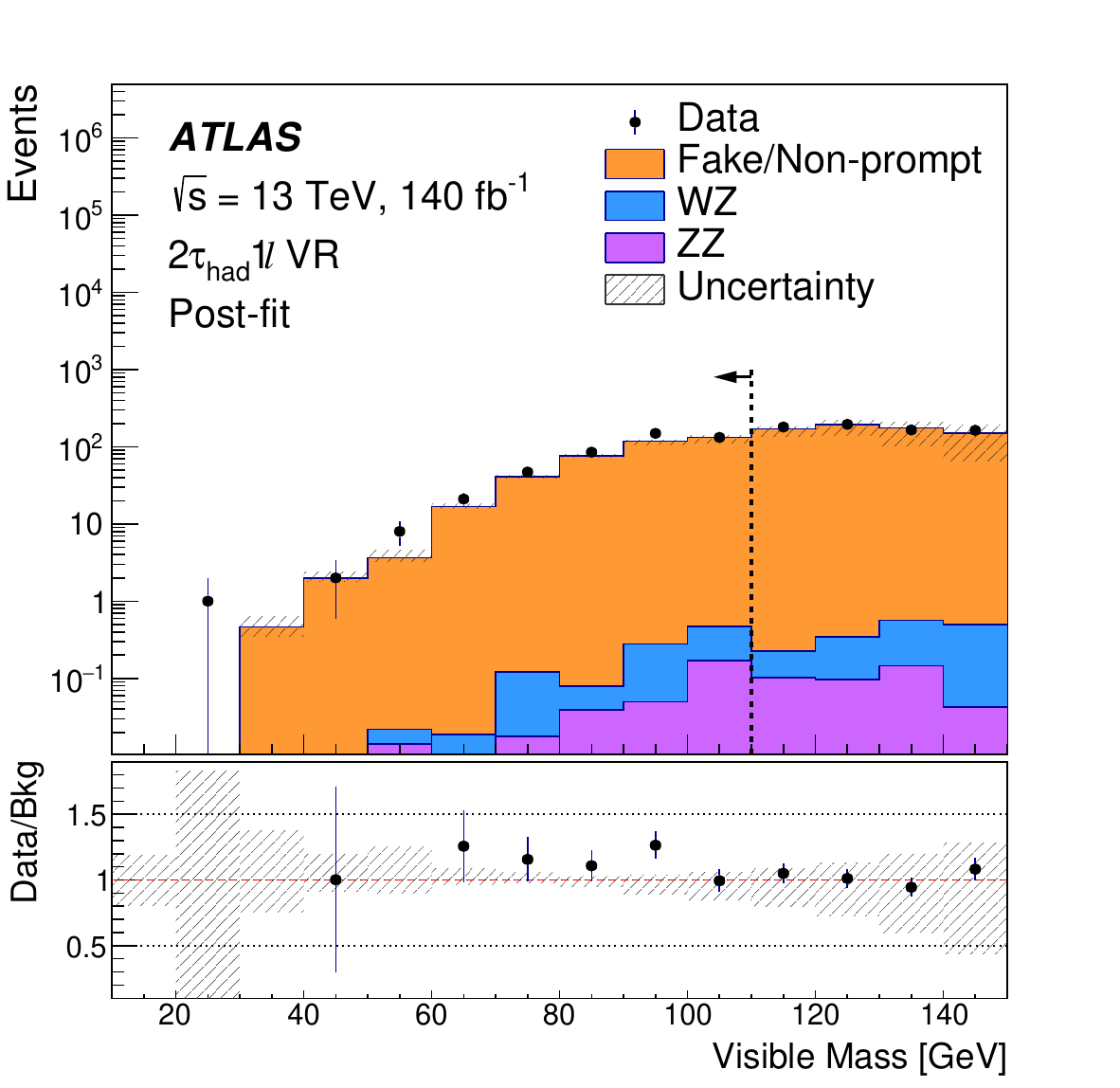}
\includegraphics[width=0.49\linewidth]{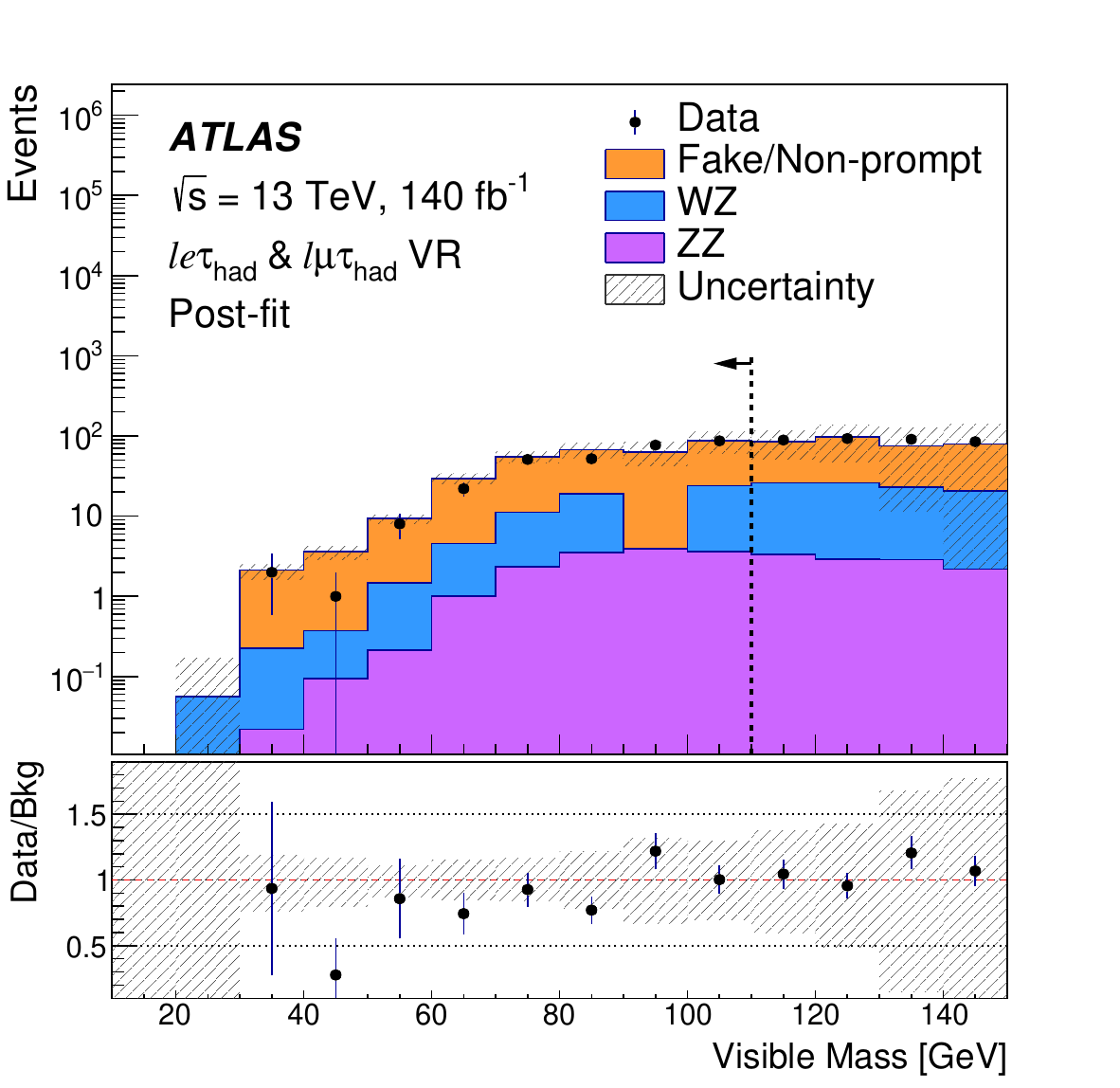}
\caption{Post-fit visible mass, $m_{\text{vis}}^{\text{3-body}}$, distribution in the $1\ell2\tau_{\text{had}}$ (left) and $\ell e\tau_{\text{had}} / \ell \mu\tau_{\text{had}}$ (right) validation regions. The data (points) are compared with the background prediction, with the post-fit total uncertainty shown as a hashed band. The vertical dashed line at $m_{\text{vis}}^{\text{3-body}} = 110\,\GeV$ indicates the upper bound of the regions. The agreement between data and prediction in the high-mass sideband ($m_{\text{vis}}^{\text{3-body}} > 110\,\GeV$) validates the modelling of the fake/non-prompt background.}
\label{fig:enter-label-1}
\end{figure}

In the profile-likelihood fit, the $\ell e\tau_{\text{had}}$ / $\ell \mu\tau_{\text{had}}$ validation regions are binned in the $\pt$ and $|\eta|$ of the subleading electron (muon). The $1\ell 2\tau_{\text{had}}$ VR is binned in the visible \pt and number of tracks of the subleading $\tau_{\text{had}}$. The binning is exactly the same as the one used for the fake-factor parameterisation. The pre-fit difference between data and expectation, which is found to be approximately 10\% across the validation regions, is incorporated as an additional non-closure uncertainty. The post-fit values are used as a refined estimate of the fake factor values to be used when estimating the fake/non-prompt background in the SRs. Figure~\ref{fig:enter-label-1} shows the post-fit comparison between data and expectation in VRs.

A statistical combination of the two signal regions is performed. Signal hypotheses for $m_a$ ranging from $15\,\GeV$ to $60\,\GeV$ are tested against the background-only hypothesis by comparing the observed data to the expected background using a profile likelihood ratio test statistic~\cite{Cowan:2010js} built from the product of Poisson densities in the two regions.
Nuisance parameters and Gaussian constraint terms~\cite{Cowan:2010js} are added to model systematic uncertainties.
Systematic uncertainties related to the electron, muon, $\tau$-lepton, and jet energy scale and resolution, and the electron, muon and $\tau$-lepton efficiencies are included~\cite{EGAM-2021-02,MUON-2018-03,ATL-PHYS-PUB-2015-045,JETM-2018-05}.
Systematic uncertainties in the modelling of the prompt $ZZ$ and $WZ$ background are estimated by varying PDFs and renormalisation and factorisation scales~\cite{Butterworth:2015oua}.
The dominant systematic uncertainties in this search are associated with the modelling of the fake/non-prompt background, in particular uncertainties from variations in the fake composition, with a magnitude of these uncertainties less than 10\%. %

\begin{figure}[htbp!]
\centering
\includegraphics[width=0.5\linewidth]{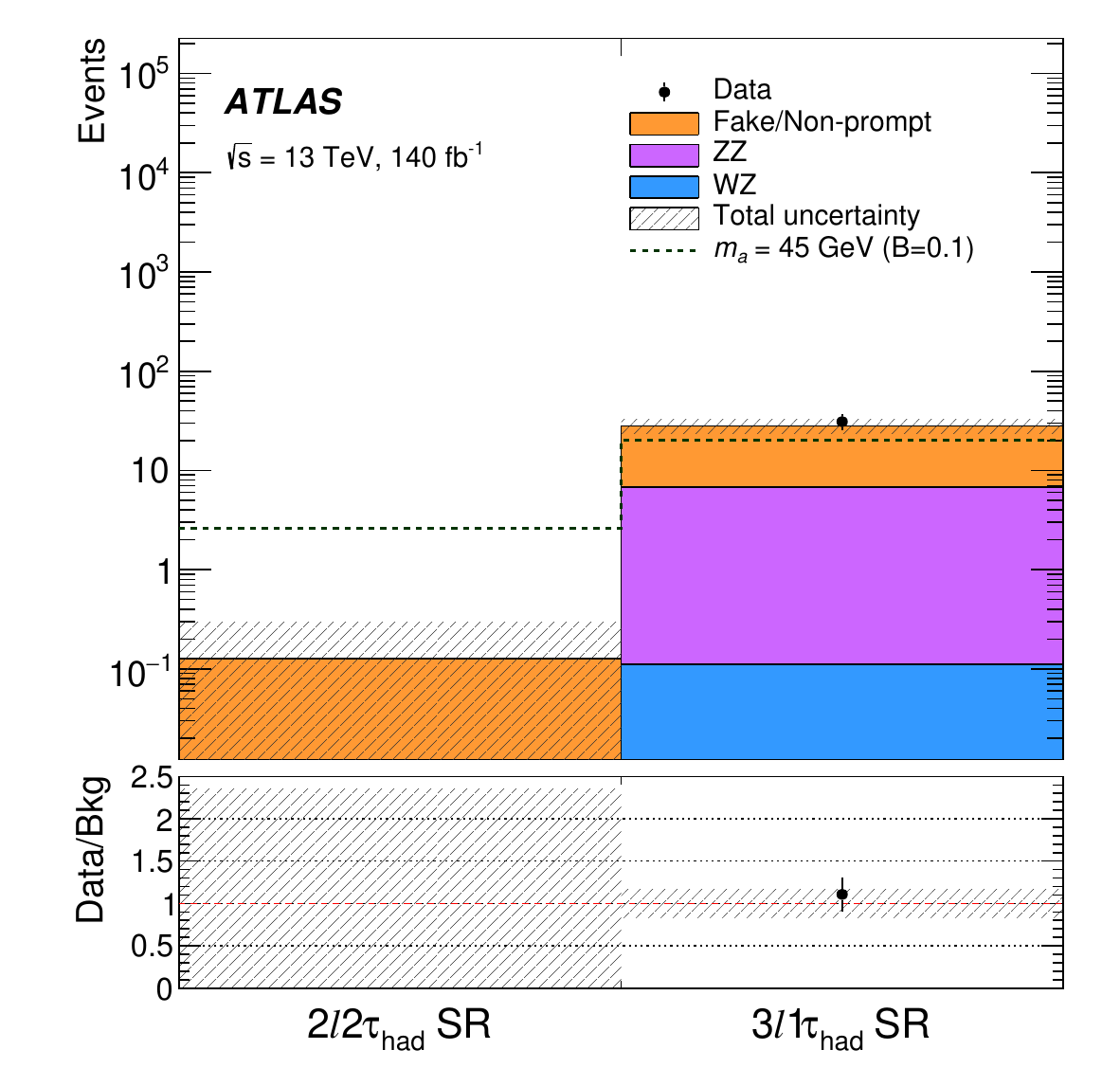}
\caption{Post-fit yield in the $2\ell2\tau_{\text{had}}$ and $3\ell1\tau_{\text{had}}$ signal regions from the background-only fit. The data (points) are compared with the background prediction, with the background total uncertainty shown as a hashed band. No data events were observed in the $2\ell2\tau_{\text{had}}$ SR.
The expected number of signal events is shown for the $m_a=45\,\GeV$ hypothesis assuming $\mathcal{B}(H\to aa\to 4\tau)=10\%$ and the inclusive Higgs boson cross-section.}
\label{fig:enter-label-2}
\end{figure}

\begin{figure}[tbp!]
\centering
\includegraphics[width=0.6\linewidth]{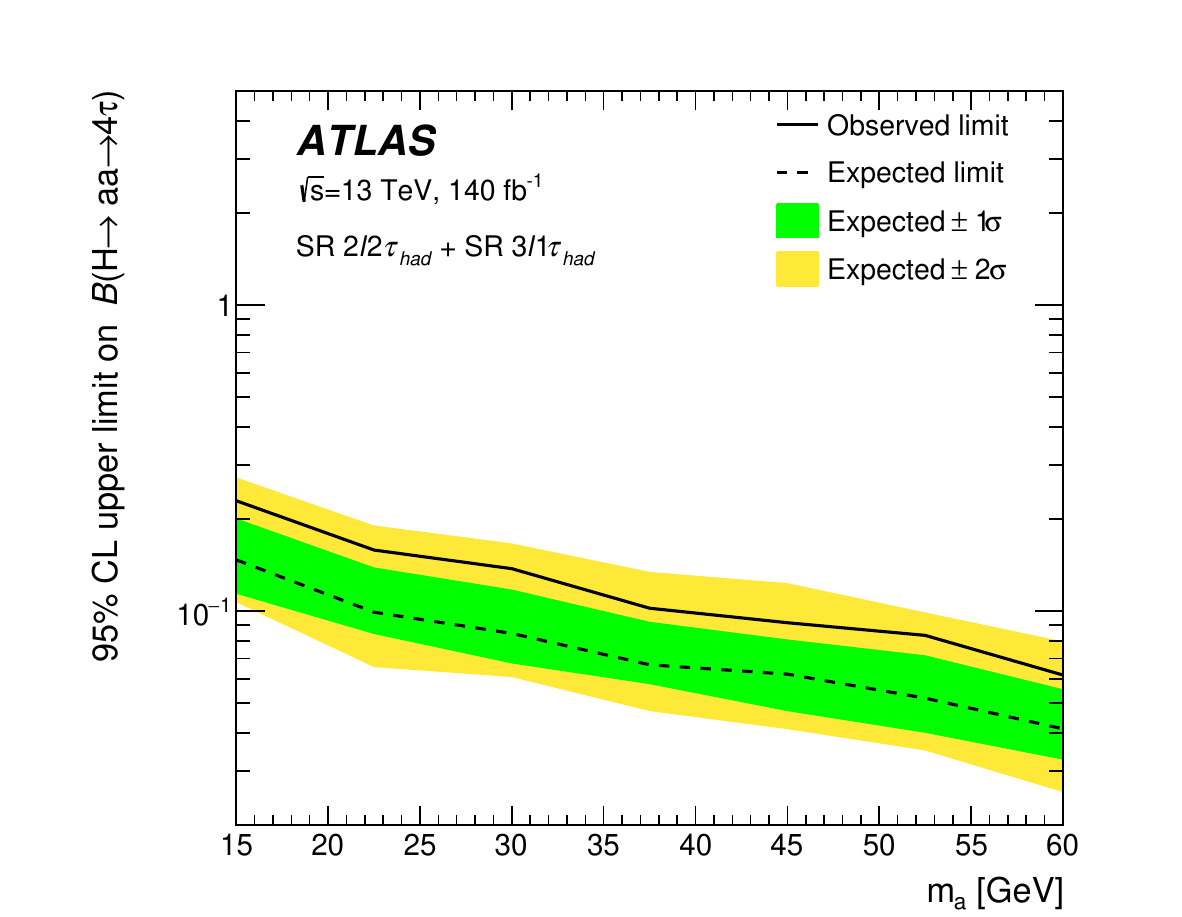}
\caption{The expected and observed 95\% CL upper limits on the branching ratio of $H \to aa \to \tau^+\tau^-\tau^+\tau^-$ in the mass range $15<m_a<60\,\GeV$. }
\label{fig:enter-label-3}
\end{figure}

Table~\ref{tab:my_label_1} and Figure~\ref{fig:enter-label-2} show the expected yields and uncertainties for the two signal regions. In the $2\ell2\tau_{\text{had}}$ SR, zero data events are observed, while 31 data events are observed in the $3\ell1\tau_{\text{had}}$ SR.
The total number of expected background events after the fit are $0.12^{+0.17}_{-0.12}$ and $28.0 \pm 4.6$.
The numbers of expected signal events for a 10\% of branching ratio range from 0.6 (10.8) to 4.3 (27.3) in the $2\ell2\tau_{\text{had}}$($3\ell1\tau_{\text{had}}$) signal region for $m_a$ from 15 to 60~\GeV.
The number of observed events in both signal regions is consistent with the expected Standard Model background estimate. Since no significant excess above the Standard Model prediction is observed, upper limits are set on the branching fraction of the Higgs boson to $aa\to \tau^+\tau^-\tau^+\tau^-$.
The limits are derived using the CLs method~\cite{Read:2002hq} with a modified test statistic $\tilde{q}_{\mu}$~\cite{Cowan:2010js} that is set to zero if the expected number of events is below the best fit value.
The upper limits are determined by generating and evaluating multiple pseudo-experiments for each signal hypothesis.
The observed (expected) 95\% CL upper limit on the branching ratio $\mathcal{B}(H \to aa \to \tau^+\tau^-\tau^+\tau^-)$ varies from 0.23 (0.15) to 0.06 (0.04) in the mass range from $m_a=15\,\GeV$ to $m_a=60\,\GeV$, as depicted in Figure~\ref{fig:enter-label-3}.
The range of the observed (expected) 95\% CL upper limit on $\mathcal{B}(H \to aa \to \tau^+\tau^-\tau^+\tau^-)$ in the $2\ell2\tau_{\text{had}}$ signal region is 0.53--0.07 (0.54--0.07).
In the $3\ell1\tau_{\text{had}}$ signal region, the corresponding observed (expected) limit is 0.26--0.10 (0.17--0.06).

\begin{table}[htbp!]
\caption{Total number of observed and expected events in the two signal regions after the background--only fit, where the signal strength is fixed at zero in the fit. %
The expected number of signal events is listed only for the $m_a=45\,\GeV$ hypothesis assuming $\mathcal{B}(H\to aa\to 4\tau)=10\%$ and the inclusive Higgs boson cross-section. The uncertainties include both statistical and systematic uncertainties. %
}
\centering
\renewcommand{\arraystretch}{1.2}
\begin{tabular}{lcc}
\toprule
\textbf{Process} & \textbf{$2\ell2\tau_{\text{had}}$ SR} &
\textbf{$3\ell1\tau_{\text{had}}$ SR} \\
\midrule
\textbf{Data}     &  0 & 31 \\
\midrule
\textbf{Total background} & $0.12^{+0.17}_{-0.12}$ & $28.0\pm 4.6$ \\
\quad\quad Fake/non-prompt & $0.12^{+0.17}_{-0.12}$ & $21.1\pm 4.7$ \\
\quad\quad $WZ$ & $<0.01$ & $0.11 \pm 0.02$\\
\quad\quad $ZZ$ & $0.01 \pm 0.01$ & $6.7\pm 0.7$\\
\midrule
\textbf{$H\to aa\to 4\tau$ ($\mathcal{B}=10\%$ $m_a=45\,\GeV$)} & $2.6 \pm 0.3$ & $20.5 \pm 2.4$ \\
\bottomrule
\end{tabular}
\label{tab:my_label_1}
\end{table}

This result constitutes the first search by the ATLAS Collaboration for the exotic Higgs boson decays $H \to aa \to \tau^+\tau^-\tau^+\tau^-$ in the mass range $15<m_a<60\,\GeV$, complementing the low mass search in the mass range of $4<m_a<15\,\GeV$ with the upper limit on $\mathcal{B}(H\rightarrow aa \rightarrow 4\tau)$ from 0.03 to 0.10~\cite{HMBS-2024-25}. The sensitivity is largely limited by the statistical uncertainty of the data. The limits on $\mathcal{B}(H \to aa \to \tau^+\tau^-\tau^+\tau^-)$ are complementary to previous searches in different final states within type-III 2HDM+s models with large values of $\tan\beta$ in the mass range $15<m_a<60\,\GeV$.

\FloatBarrier

\section*{Acknowledgements}
%

%

% The next lines are included from the .//acknowledgements/Acknowledgements.tex input file
%
%

%
%

We thank CERN for the very successful operation of the LHC and its injectors, as well as the support staff at
CERN and at our institutions worldwide without whom ATLAS could not be operated efficiently.

The crucial computing support from all WLCG partners is acknowledged gratefully, in particular from CERN, the ATLAS Tier-1 facilities at TRIUMF/SFU (Canada), NDGF (Denmark, Norway, Sweden), CC-IN2P3 (France), KIT/GridKA (Germany), INFN-CNAF (Italy), NL-T1 (Netherlands), PIC (Spain), RAL (UK) and BNL (USA), the Tier-2 facilities worldwide and large non-WLCG resource providers. Major contributors of computing resources are listed in Ref.~\cite{ATL-SOFT-PUB-2026-001}.

We gratefully acknowledge the support of ANPCyT, Argentina; YerPhI, Armenia; ARC, Australia; BMWFW and FWF, Austria; ANAS, Azerbaijan; CNPq and FAPESP, Brazil; NSERC, NRC and CFI, Canada; CERN; ANID, Chile; CAS, MOST and NSFC, China; Minciencias, Colombia; MEYS CR, Czech Republic; DNRF and DNSRC, Denmark; IN2P3-CNRS and CEA-DRF/IRFU, France; SRNSFG, Georgia; BMFTR, HGF and MPG, Germany; GSRI, Greece; RGC and Hong Kong SAR, China; ICHEP and Academy of Sciences and Humanities, Israel; INFN, Italy; MEXT and JSPS, Japan; CNRST, Morocco; NWO, Netherlands; RCN, Norway; MNiSW, Poland; FCT, Portugal; MNE/IFA, Romania; MSTDI, Serbia; MSSR, Slovakia; ARIS and MVZI, Slovenia; DSI/NRF, South Africa; MICIU/AEI, Spain; SRC and Wallenberg Foundation, Sweden; SERI, SNSF and Cantons of Bern and Geneva, Switzerland; NSTC, Taipei; TENMAK, T\"urkiye; STFC/UKRI, United Kingdom; DOE and NSF, United States of America.

Individual groups and members have received support from BCKDF, CANARIE, CRC and DRAC, Canada; CERN-CZ, FORTE and PRIMUS, Czech Republic; COST, ERC, ERDF, Horizon 2020 and Marie Sk{\l}odowska-Curie Actions, European Union; Investissements d'Avenir Labex, Investissements d'Avenir Idex and ANR, France; DFG and AvH Foundation, Germany; Herakleitos, Thales and Aristeia programmes co-financed by EU-ESF and the Greek NSRF, Greece; BSF-NSF and MINERVA, Israel; NCN and NAWA, Poland; La Caixa Banking Foundation, CERCA and AGAUR programs from Generalitat de Catalunya and PROMETEO and GenT Programmes Generalitat Valenciana, Spain; G\"{o}ran Gustafssons Stiftelse, Sweden; The Royal Society and Leverhulme Trust, United Kingdom; Eric and Wendy Schmidt Fund for Strategic Innovation, United States of America.

In addition, individual members wish to acknowledge support from Chile: Agencia Nacional de Investigaci\'on y Desarrollo (ANID FONDECYT reg. 1230987, FONDECYT 1230812, FONDECYT 1240864, Fondecyt 3240661, Fondecyt Regular 1240721); China: Chinese Ministry of Science and Technology (MOST-2023YFA1605700, MOST-2023YFA1609300), National Natural Science Foundation of China (NSFC - 12175119, NSFC 12275265); Czech Republic: Czech Science Foundation (GACR - 24-11373S), Ministry of Education Youth and Sports (ERC-CZ-LL2327, FORTE CZ.02.01.01/00/22\_008/0004632), PRIMUS Research Programme (PRIMUS/21/SCI/017); EU: H2020 European Research Council (ERC - 101002463); European Union: European Research Council (BARD No. 101116429, ERC - 948254, ERC 101089007), European Regional Development Fund (HE COFUND GA No.101081355, ERDF), European Union, Future Artificial Intelligence Research (FAIR-NextGenerationEU PE00000013), Marie Sklodowska-Curie Actions (GAP-101168829); France: Agence Nationale de la Recherche (ANR-21-CE31-0013, ANR-21-CE31-0022, ANR-22-EDIR-0002, ANR-24-CE31-0504-01); Germany: Deutsche Forschungsgemeinschaft (DFG - 469666862); China: Research Grants Council (GRF); Italy: Ministero dell'Università e della Ricerca (NextGenEU 153D23001490006 M4C2.1.1, NextGenEU I53D23000820006 M4C2.1.1, NextGenEU I53D23001490006 M4C2.1.1, SOE2024\_0000023); Japan: Japan Society for the Promotion of Science (JSPS KAKENHI  JP25H0063, JSPS KAKENHI JP22H01227, JSPS KAKENHI JP22H04944, JSPS KAKENHI JP22KK0227, JSPS KAKENHI JP24K23939, JSPS KAKENHI JP24KK0251, JSPS KAKENHI JP25H00650, JSPS KAKENHI JP25H01291, JSPS KAKENHI JP25K01011, JSPS KAKENHI JP25K01023); Norway: Research Council of Norway (RCN-314472); Poland: Ministry of Science and Higher Education (IDUB AGH, POB8, D4 no 9722), Polish National Science Centre (NCN 2021/42/E/ST2/00350, NCN OPUS 2023/51/B/ST2/02507, NCN OPUS nr 2022/47/B/ST2/03059, NCN UMO-2019/34/E/ST2/00393, UMO-2022/47/O/ST2/00148, UMO-2023/49/B/ST2/04085, UMO-2023/51/B/ST2/00920, UMO-2024/53/N/ST2/00869); Spain: Agència de Gestió d'Ajuts Universitaris i de Recerca. (AGAUR - 2023 BP 00141), Generalitat Valenciana (ASFAE/2022/008), Ministry of Science and Innovation (RYC2019-028510-I, RYC2020-030254-I, RYC2021-031273-I, RYC2022-038164-I), Ministerio de Ciencia, Innovación y Universidades/Agencia Estatal de Investigaci\'on (EU NextGenerationEU (PRTR-C17.I1), PID2022-142604OB-C22); Sweden: Carl Trygger Foundation (Carl Trygger Foundation CTS 22:2312), Swedish Research Council (Swedish Research Council 2023-04654, VR 2021-03651, VR 2022-03845, VR 2022-04683, VR 2023-03403, VR 2024-05451), Knut and Alice Wallenberg Foundation (KAW 2018.0458, KAW 2023.0366); Switzerland: Swiss National Science Foundation (SNSF - PCEFP2\_194658); United Kingdom: The Binks Trust, Royal Society (NIF-R1-231091); United States of America: U.S. Department of Energy (ECA DE-AC02-76SF00515), John Templeton Foundation (John Templeton Foundation 63206), Neubauer Family Foundation.

%
%

% End of text imported from the .//acknowledgements/Acknowledgements.tex input file

%
\clearpage
\printbibliography
\clearpage
\input{atlas_authlist}
%

%
%

%

%
%
%
%

\end{document}

%% file: atlas_authlist.tex
% ATLAS Collaboration author list
% Reference date of HMBS-2024-35 is 2025-12-02
% Author list last updated on date 23-APR-26
% Data extracted on 23-Apr-2026 for paper reference HMBS-2024-35
% at 7:01pm
 
\begin{flushleft}
\hypersetup{urlcolor=black}
{\Large The ATLAS Collaboration}

\bigskip

\AtlasOrcid[0000-0002-6665-4934]{G.~Aad}$^\textrm{\scriptsize 102}$,
\AtlasOrcid[0000-0001-7616-1554]{E.~Aakvaag}$^\textrm{\scriptsize 17}$,
\AtlasOrcid[0000-0002-5888-2734]{B.~Abbott}$^\textrm{\scriptsize 121}$,
\AtlasOrcid[0000-0002-0287-5869]{S.~Abdelhameed}$^\textrm{\scriptsize 83b}$,
\AtlasOrcid[0000-0002-1002-1652]{K.~Abeling}$^\textrm{\scriptsize 54}$,
\AtlasOrcid[0000-0001-5763-2760]{N.J.~Abicht}$^\textrm{\scriptsize 48}$,
\AtlasOrcid[0000-0002-8496-9294]{S.H.~Abidi}$^\textrm{\scriptsize 30}$,
\AtlasOrcid[0009-0003-6578-220X]{M.~Aboelela}$^\textrm{\scriptsize 44}$,
\AtlasOrcid[0000-0002-9987-2292]{A.~Aboulhorma}$^\textrm{\scriptsize 36e}$,
\AtlasOrcid[0000-0001-5329-6640]{H.~Abramowicz}$^\textrm{\scriptsize 154}$
% ...
\bigskip
\\

% ...
$^{17}$Department for Physics and Technology, University of Bergen, Bergen; Norway.\\
% ...
$^{30}$Physics Department, Brookhaven National Laboratory, Upton NY; United States of America.\\
% ...
$^{36}$$^{(a)}$Facult\'e des Sciences Ain Chock, Universit\'e Hassan II de Casablanca;$^{(b)}$Facult\'{e} des Sciences, Universit\'{e} Ibn-Tofail, K\'{e}nitra;$^{(c)}$Facult\'e des Sciences Semlalia, Universit\'e Cadi Ayyad, LPHEA-Marrakech;$^{(d)}$LPMR, Facult\'e des Sciences, Universit\'e Mohamed Premier, Oujda;$^{(e)}$Facult\'e des sciences, Universit\'e Mohammed V, Rabat;$^{(f)}$Institute of Applied Physics, Mohammed VI Polytechnic University, Ben Guerir; Morocco.\\
% ...
$^{44}$Physics Department, Southern Methodist University, Dallas TX; United States of America.\\
% ...
$^{48}$Fakult\"{a}t Physik, Technische Universit{\"a}t Dortmund, Dortmund; Germany.\\
% ...
$^{54}$II. Physikalisches Institut, Georg-August-Universit\"{a}t G\"ottingen, G\"ottingen; Germany.\\
% ...
$^{83}$$^{(a)}$Khalifa University of Science and Technology, Abu Dhabi;$^{(b)}$New York University Abu Dhabi, Abu Dhabi;$^{(c)}$United Arab Emirates University, Al Ain;$^{(d)}$University of Sharjah, Sharjah; United Arab Emirates.\\
% ...
$^{102}$CPPM, Aix-Marseille Universit\'e, CNRS/IN2P3, Marseille; France.\\
% ...
$^{121}$Homer L. Dodge Department of Physics and Astronomy, University of Oklahoma, Norman OK; United States of America.\\
% ...
$^{154}$Raymond and Beverly Sackler School of Physics and Astronomy, Tel Aviv University, Tel Aviv; Israel.\\
% ...
$^{a}$ Also at Affiliated with an institute formerly covered by a cooperation agreement with CERN.\\
$^{b}$ Also at An-Najah National University, Nablus; Palestine.\\
$^{c}$ Also at Borough of Manhattan Community College, City University of New York, New York NY; United States of America.\\
$^{d}$ Also at Center for Interdisciplinary Research and Innovation (CIRI-AUTH), Thessaloniki; Greece.\\
$^{e}$ Also at Centre of Physics of the Universities of Minho and Porto (CF-UM-UP); Portugal.\\
$^{f}$ Also at CERN, Geneva; Switzerland.\\
% ...

\end{flushleft}

% Created with Glance <Atlas.Glance@cern.ch>